\documentclass{article}
\usepackage{amsmath,graphicx,mlspconf,multicol,graphicx,multirow,mathtools,cite,balance}% \usepackage{graphicx}
\usepackage[table,xcdraw]{xcolor}

\copyrightnotice{U.S.\ Government work not protected by U.S.\ copyright}

\def\cheat{{\vspace{-.8ex}}}

\title{WEnets: A Convolutional Framework for Evaluating Audio Waveforms}

\name{Andrew A. Catellier and Stephen D. Voran}
\address{Institute for Telecommunication Sciences, 325 Broadway, Boulder, Colorado, USA \\
         \{acatellier, svoran\}@ntia.gov}

\begin{document}
%\ninept
%

\maketitle
\begin{abstract}
We describe a new convolutional framework for waveform evaluation, WEnets, and build a Narrowband Audio Waveform Evaluation Network, or NAWEnet, using this framework.
NAWEnet is single-ended (or no-reference) and was trained three separate times in order to emulate PESQ, POLQA, or STOI with testing correlations 0.95, 0.92, and 0.95, respectively when training on only 50\% of available data and testing on 40\%.
Stacks of 1-D convolutional layers and non-linear downsampling learn which features are important for quality or intelligibility estimation.
This straightforward architecture simplifies the interpretation of its inner workings and paves the way for future investigations into higher sample rates and accurate no-reference subjective speech quality predictions.
\end{abstract}
\begin{keywords}
CNN, neural nets, no reference, single ended, speech intelligibility, speech quality
\end{keywords}
\section{Introduction}
\label{sec:intro}

Measuring quality and other abstract properties of audio signals is essential to the development, deployment, maintenance, and marketing of audio-related products and services.
In telecommunications, speech quality and intelligibility are critical to customer satisfaction and thus to successful products and services.
Perception-based measurements have evolved to properly account for
the various distortions produced as digital encoding, transmission, and decoding have permeated telecommunications networks and equipment. Earlier work is summarized in \cite{VoranChapter} and current popular measurements include Perceptual Evaluation of Speech Quality (PESQ) \cite{PESQ}, Perceptual Objective Listening Quality Analysis (POLQA) \cite{POLQA}, and Short-Time Objective Intelligibility Measure (STOI) \cite{STOI}.

Measurement algorithms typically transform transmitted and received speech signals into perceptual domain representations (emulating hearing) and then compare those two representations (emulating listening, attention, and judgment).
A common goal is that results should agree with those produced by human subjects in formal, controlled, speech quality or speech intelligibility experiments. 
Thus the measurements are often viewed as ``estimators'' of the ``true'' values from experiments.
``Full-reference'' (FR) estimators have produced impressive and useful results but only when the transmitted and received speech signals are both available for evaluation.
PESQ and POLQA are prominent examples of FR perception-based speech quality estimators and STOI is an effective FR perception-based speech intelligibility estimator.

``No-reference'' (NR) (also called ``non-intrusive'') approaches offer the ability to estimate using only the received speech.
This capability can provide significant additional opportunities, including live monitoring, fault detection, or optimization of telecommunications systems.

Broadly speaking, much of the NR speech estimation work (e.g., \cite{RFK1994, P.563, ANIQUE+, NR-STOI}) has been driven by models for clean and distorted speech along with a means for analyzing received speech and properly locating it within the space defined by those models.
As machine learning (ML) tools became more developed, powerful, and available, they were naturally incorporated into NR speech evaluation algorithms \cite{Falk2006, Soni2016, Fu, Salehi2018, Spille2018, Huber2018, Pardo2018, Avila2019, Mittag2019, Ooster2019}.
Algorithms typically start with the extraction of known-relevant features (e.g. magnitude spectrogram, Mel-spectral or Mel-cepstral features, pitch values, voice activity) from the received speech.
This is followed by application of assorted ML structures to learn and codify the mapping between these features and some target quantity relating to the suitability of speech (e.g., quality, intelligibility, listening effort).
In some cases automatic speech recognition inspired modeling is invoked as well.  

These approaches are certainly well-motivated.
Extraction of features compresses speech representations (for efficiency) while retaining information relevant to establishing accurate mappings to a target.
But starting with established features does constrain the solution space accordingly.
Given the power of convolutional neural networks (CNN) it is now possible to eliminate any assumptions or restrictions explicit or implicit in feature extraction and allow a CNN to operate directly on speech waveforms, in effect building the best features for solving the problem at hand.

In this work we require the machine to learn which features are appropriate for a waveform evaluation task.
We establish a framework named Waveform Evaluation networks (WEnets) and then demonstrate the value of this framework by developing a Narrowband Audio Waveform Evaluation Network (NAWEnet) that performs NR prediction of NB speech quality or intelligibility.
We describe a method to generate training, testing, and validation data for this task.
Using 133.5 hours of training data and 106.8 hours of testing data we achieve per-segment prediction-to-target correlations ($\rho_{seg}$) above 0.91.
Due to the straightforward architecture of NAWEnet we expect that our future work may provide interpretation of its inner workings.
We also expect to extend the approach to address the evaluation of wideband or fullband speech or even music.

\section{Network Design}
\label{sec:design}
\subsection{WEnets Framework Principles}
\label{ssec:wenet_principles}

CNN architectures trained to find objects in images are often required to learn how to perform a task regardless of the scale (or spatial sample rate) of the object to be recognized.
In a typical image classification database like ImageNet, objects of any given class can be found with varying spatial sampling rates.
In order for any ML process to successfully classify objects in images, the ML process must learn to find the object at many different spatial sampling rates.

VGG, a CNN architecture developed for image classification purposes by the Visual Geometry Group \cite{DBLP:journals/corr/SimonyanZ14a} is composed of sections containing one or more sequential convolutional layers, a non-linear activation (rectified linear unit, or ReLU, in this case), and a max-pooling layer.
Each max-pooling layer essentially downsamples representations of the input image thus enabling the next section to operate at a higher level of abstraction.
These sections are stacked until the image has been sufficiently downsampled such that all $f_n$ representations---512 in the case of VGG---can feasibly be used as an input to a classification dense network.
This architecture is proven to find objects at multiple spatial sample rates.

Like VGG, WEnets are composed of a CNN that is used to extract features and a dense network that computes a target speech quality or intelligibility estimate using the extracted features.

Unlike the images found in ImageNet, audio (and many other one-dimensional) signals have a fixed sample rate measured in units of time rather than units of distance.
If operating on audio, a convolutional architecture as described above will not need to find a fixed-duration feature (or object) at multiple timescales.
That is, a feature that is $x$ seconds in length will always be $x\times f_s$ samples long where $f_s$ is the sample rate of the audio signal.

Due to this property, waveform-specific CNN architectures need not use network depth as a method to robustly handle scale/sample rate variance.
Rather, the depth of waveform-specific architectures can be informed by the desired input sample rate and the time-scale of the features to be extracted.
We hypothesize that stacked convolutional layers can be used to find waveform features and waveform distortions with time durations consistent with the input sample rates ($\hat{f_s}$) of the layers.
Thus the WEnets framework uses stacked convolutional layers, non-linear activations, and pooling to extract and process the information necessary to evaluate waveforms.

\begin{table}[tbp]
\centering
\begin{tabular}{c|crrcr}
$S$                    & layer type & \multicolumn{1}{c}{$\hat{f_s}$ (Hz)} & \multicolumn{1}{c}{$l_{in}$} & $s_l$ (ms)             & \multicolumn{1}{c}{$l_{out}$} \\ \hline
\multirow{4}{*}{$S_1$} & C-192-11   & \multirow{4}{*}{8,000}               & \multirow{4}{*}{24,000}      & \multirow{4}{*}{0.125} & \multirow{4}{*}{6,000}        \\
                       & B          &                                      &                              &                        &                               \\
                       & P-192      &                                      &                              &                        &                               \\
                       & A-4        &                                      &                              &                        &                               \\ \hline
\multirow{4}{*}{$S_2$} & C-192-7    & \multirow{4}{*}{2,000}               & \multirow{4}{*}{6,000}       & \multirow{4}{*}{0.5}   & \multirow{4}{*}{3,000}        \\
                       & B          &                                      &                              &                        &                               \\
                       & P-192      &                                      &                              &                        &                               \\
                       & M-2        &                                      &                              &                        &                               \\ \hline
\multirow{4}{*}{$S_3$} & C-256-7    & \multirow{4}{*}{1,000}               & \multirow{4}{*}{3,000}       & \multirow{4}{*}{1}     & \multirow{4}{*}{750}          \\
                       & B          &                                      &                              &                        &                               \\
                       & P-256      &                                      &                              &                        &                               \\
                       & M-4        &                                      &                              &                        &                               \\ \hline
\multirow{5}{*}{$S_4$} & C-512-7    & \multirow{5}{*}{250}                 & \multirow{5}{*}{750}         & \multirow{5}{*}{4}     & \multirow{5}{*}{250}          \\
                       & C-512-7    &                                      &                              &                        &                               \\
                       & B          &                                      &                              &                        &                               \\
                       & P-512      &                                      &                              &                        &                               \\
                       & M-3        &                                      &                              &                        &                               \\ \hline
\multirow{5}{*}{$S_5$} & C-512-7    & \multirow{5}{*}{83.3}                & \multirow{5}{*}{250}         & \multirow{5}{*}{12}    & \multirow{5}{*}{125}          \\
                       & C-512-7    &                                      &                              &                        &                               \\
                       & B          &                                      &                              &                        &                               \\
                       & P-512      &                                      &                              &                        &                               \\
                       & M-2        &                                      &                              &                        &                              
\end{tabular}
\caption{NAWEnet convolutional architecture: each section $S_n$ contains one or two 1D convolution layers C-$f_n$-$f_l$ with $f_n$ filters and filter length $f_l$, stride of 1, and zero padding equal to $\lfloor\frac{f_l}2\rfloor$. B denotes a batch normalization layer. P-$f_n$ indicates a PReLU activation with $f_n$ parameters. A-$k$ is an average pooling layer with pooling kernel size $k$; M-$k$ a max pooling layer. Number of input and output samples are given by $l_{in}$ and $l_{out}$, effective sample rate by $\hat{f_s}$, and effective sample spacing by $s_l$.} 
\label{table:conv_arch}
\end{table}

\begin{table}[tbp]
\centering
\begin{tabular}{c|rr}
\multicolumn{1}{l|}{$L$} & \multicolumn{1}{c}{$d_i$} & \multicolumn{1}{c}{$d_o$} \\ \hline
$L_1$                     & 64,000                    & 512                       \\
$L_2$                     & 512                       & 512                       \\
$L_3$                     & 512                       & 1                        
\end{tabular}
\caption{NAWEnet dense net architecture: each layer ($L$) has $d_i$ inputs and $d_o$ outputs. Dense layers 1 and 2 are followed by PReLU with $d_o$ parameters and a dropout layer with $p=0.5$.}
\label{table:dense_arch}
\end{table}

\subsection{NAWEnet Implementation}
\label{ssec:nawenet_implementation}

The architecture for the NAWEnet, a narrowband-audio implementation of the WEnets framework, is shown in Tables \ref{table:conv_arch} and \ref{table:dense_arch}.
We designed the CNN feature extractor with speech and speech coding in mind.
The $\hat{f_s}$ at the input to $S_1$ is the norm for NB speech and preserves all waveform details. 
The $\hat{f_s}$ going into $S_2$-$S_4$ support the range of frequencies where the lower formants of human speech can be found.
In $S_5$ $s_l$ is 12 ms, which is on the order of the length of speech coding frames and the length of packets used to transmit voice over the internet.

In each convolutional section, the network learns $f_n$ representations of the input signal, $f_n$ batch normalization \cite{DBLP:journals/corr/IoffeS15} parameters, and $f_n$ parameters that control the slope for the $x < 0$ portion of the Parametric ReLU (PReLU) \cite{DBLP:journals/corr/HeZR015} activation function.
Each convolutional section concludes with a pooling layer where $f_n$ representations are essentially downsampled into $l_{out}$ subsamples.

In the $S_1$, average-pooling behaves somewhat like a typical downsampling process and gathers information from $k=4$ samples into one subsample.
But in subsequent sections max-pooling chooses the subsample with the highest value for input to the next section.
The combination of convolutional filtering and max-pooling coalesces relevant information and begins to form features.
As training progresses this process ultimately allows the net to learn which kinds of features are required to predict a target metric.

The final max-pooling layer in $S_5$ subsamples each of $f_n = 512$ representations to a length $l_{out}=125$ subsamples.
The output of the feature extractor is then flattened resulting in $512\times 125=64{,}000$ inputs to the dense network.
After the first two dense layers we implement dropout \cite{DBLP:journals/corr/abs-1207-0580} to minimize over-fitting.
Weights for convolutional and dense layers are initialized using the fan-out variant of the Kaiming normal method \cite{DBLP:journals/corr/HeZR015}.

Like VGG, NAWEnet requires an input of a specific size.
The inputs to NAWEnet are 3 second-long (sample rate $f_s = 8000$ smp/s) audio segments normalized to 26 dB below clipping points of $[-1, 1]$.
The choice of three seconds was driven by the active speech content in the speech files commonly used for telecommunications testing.
A target PESQ, POLQA, and STOI value is calculated for each segment.
This allows us to train NAWEnet to suit each target.

\section{Data Corpora}
\label{sec:data}
Data is essential to any ML effort and NAWEnet is no exception.  We collected and created a large number of speech recordings with a wide range of distortion types and levels.
Some of these recordings were made in our lab and in other telecommunications labs over the past decades in order to test specific telecommunications scenarios or ``conditions.''
In either case, the original undistorted speech recordings are studio-grade with very low noise and minimal reverberation and are either unfiltered or prefiltered using bandpass, intermediate reference system (IRS), or modified IRS \cite{UGST_STL} methods.
The original speech recordings were passed through telecommunications hardware or software resulting in various conditions of interest.  Then three-second segments were extracted and associated FR quality (PESQ and POLQA) and intelligibility (STOI) targets were computed for each segment. 

\begin{table}[h]
\centering
\begin{tabular}{c|ccccc}
    & original (h)& distorted (h) & language     & cond. \\ \hline
1  & 3.2 & 3.2   & NAE      & $C_a$ \\
2  & 1.8 & 1.8   & NAE      & $C_b$ \\
3  & 1.0 & 1.0   & NAE      & $C_c$  \\
4  & 2.0 & 2.0   & NAE      & $C_d$ \\
5  & 1.2 & 1.2   & Italian  & $C_d$  \\
6  & 1.8 & 1.8   & Japanese & $C_d$  \\
7  & 2.5 & 40.8  & Mixed  & $C_e$   \\
8  & 2.5 & 61.7  & Mixed    & $C_f$ \\
9   & 3.6 & 10.0    & NAE      & $C_e$   \\
10  & 3.6 & 10.0    & NAE      & $C_f$  \\
\end{tabular}
\caption{Source dataset descriptions, including duration of original speech and distorted speech in hours.}
\label{table:datasets}
\end{table}

\begin{table}[h]
\centering
\begin{tabular}{c|cl}
    & rates (kbps) & conditions \\ \hline
$C_a$  & 4.8-32 & G.728, G.726, GSM, VSELP,  \\ 
& &  IMBE, proprietary codecs, MNRU    \\
$C_b$  & 8-16   & 9 CELP variants, frame erasures \\
& & MNRU \\
$C_c$  & 2.4-64 & variable rate CELP, PCM, \\
& & analog FM, MNRU \\
$C_d$  & 16-64  & PCM, ADPCM, G.728 candidates, \\
& & MNRU \\
$C_e$  & 1.2-80 &  AMR, EVS, PCM, ADPCM,  \\
& & G.728, G.729, G.723.1, GSM,  \\
& & AMBE, MELP, proprietary codecs  \\
$C_f$  & 1.2-80 &  as in $C_e$ plus frame erasures \\
& & and concealment\cite{RORPP}, 0-25\%, \\
& & indep. and bursty, 20 ms frames\\
\end{tabular}
\caption{Summary of conditions. MNRU indicates Modulated noise reference unit\cite{MNRU}.  Level variations and tandems also included.}
\label{table:conditions}
\end{table}

Table \ref{table:datasets} summarizes attributes of the dataset contents.
``NAE'' indicates North American English. ``Mixed'' includes NAE, British English, Hindi, French, Mandarin, Finnish, German, Italian, and Japanese.
In the aggregate of the 10 datasets 86\% of the speech is NAE, 4\% is British English, and 3\% is Hindi.  
Japanese and French each account for 2\%,  Mandarin and Italian 1\% each, while Finnish and German each provide about 0.5\%. In the case of the English language, the speech content is comprised of Harvard sentences \cite{harvard_sent}. 

Source datasets 1-6 were originally created for subjective testing of specific conditions of interest.
The NAWEnet design requires three-second segments of speech.
We used software to select as many unique segments as possible, subject to a minimum speech activity factor of 75\%.
For datasets 1-6 these segments were taken from the previously produced distorted speech recordings.
Additional details regarding the conditions in each of the datasets are given in Table \ref{table:conditions}.

\begin{figure*}[ht]
  \includegraphics[width=\textwidth]{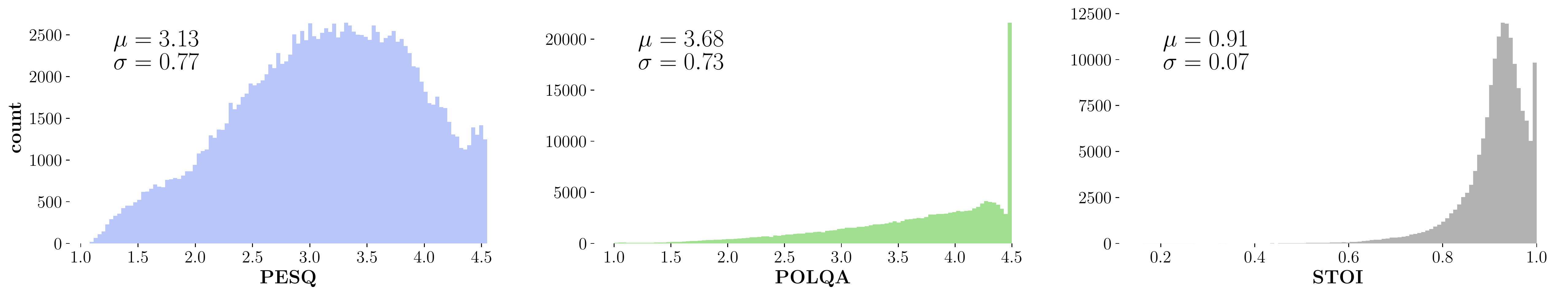}
  \caption{Histograms, means, and standard deviations of targets over all available data.\label{figure:histograms}}
\end{figure*}

Source datasets 7 and 8 were created specifically for training a NAWEnet.
They use some original undistorted speech recordings from datasets 1-6 along with recordings from the ITU-T P.501 \cite{P.501}, P-Series Supplement 23 \cite{Psup23}, and Open Speech Repository \cite{OSR} databases.
To augment existing data we allowed the segment selection software to make multiple passes through the available original speech recordings.
We required a minimum speech activity factor of 35\% in dataset 7 and 75\% in dataset 8.
When a 3 second segment with suitable speech activity was located a uniformly distributed time offset (0 to 250 ms) was applied.
These time offsets prevented any given original speech segment from appearing more than once in the datasets.
These recordings were passed through software implementations of various speech coding and transmission conditions as summarized in Table \ref{table:conditions}.

Source datasets 9 and 10 parallel 7 and 8, respectively, but the original speech recordings are exclusively from the McGill University TSP \cite{TSP} database. Thus these databases have otherwise unseen talkers and waveforms.
The minimum speech activity factor here is 43\%.

Across the ten datasets the speech-activity factor for the segments ranged from 35\% to 100\% with a mean value of 81\%.
Together the datasets include 9 languages, 148 unique talkers, over 75 different sources of distortion, and 133.5 hours of speech.

\begin{figure*}[h!]
  \includegraphics[width=\textwidth]{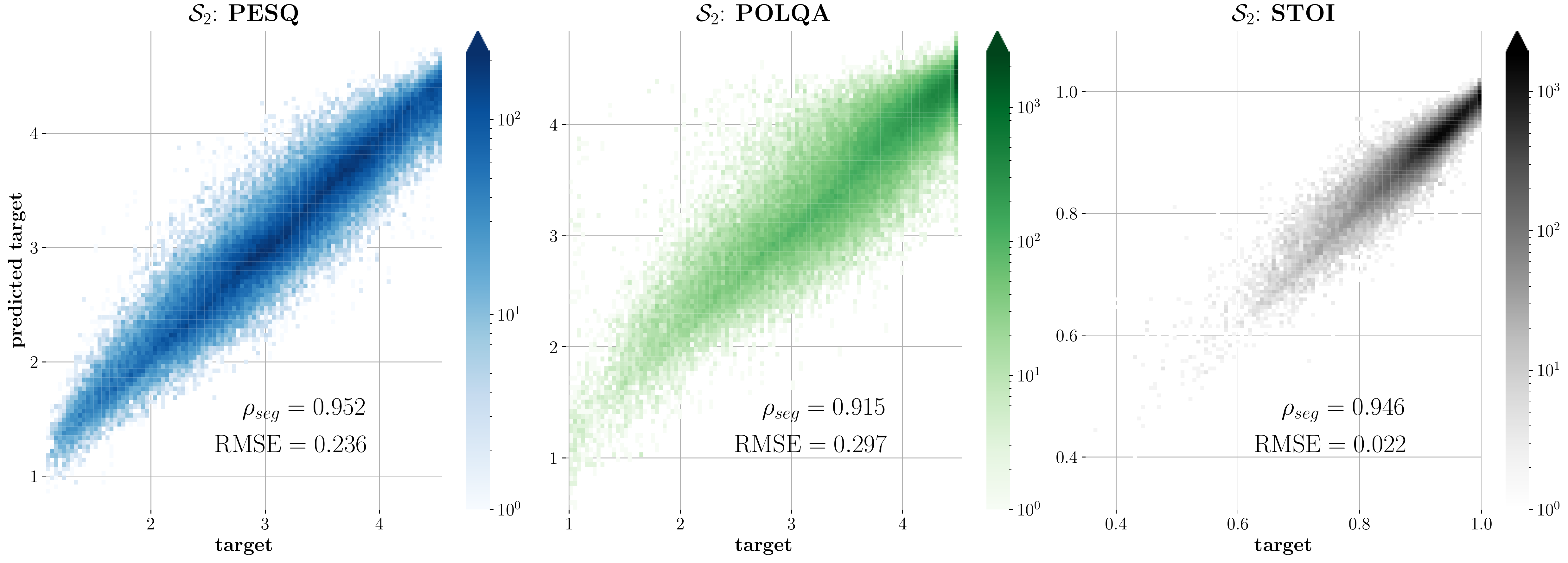}
  \caption{Two-dimensional histograms showing target vs. predicted values for PESQ, POLQA, and STOI when evaluated on the test data portion of $\mathcal{S}_2$. Segments per bin is given by the scale at the right.\label{figure:correlation}}
\end{figure*}

We used the FR estimators PESQ, POLQA, and STOI to generate three target values for each speech segment.
These targets were used for NAWEnet training, validation, and testing.
These FR metrics have limitations and may not be ideally suited for the three-second format used here but each still produces meaningful results in this application. 
In Fig. \ref{figure:histograms} we show histograms for the three target metrics over all available data.

\section{Training Methodology}
\label{sec:training}
In order to train NAWEnet we made training, validation, and testing datasets from each of the 10 available source datasets.
We built the training set by randomly selecting 50\% of the segments in a source dataset without replacement.
From the remaining segments in the source dataset we built the test set by randomly selecting 40\% of the total data without replacement.
The remaining 10\% of segments were then used for validation purposes.
Once training, testing, and validation sets had been created for each source dataset, all training sets were concatenated into one aggregated training dataset; testing and validation sets were combined in the same manner.

A constant phase inversion is inaudible so quality and intelligibility values and estimates are unchanged by phase inversion.
We wanted the networks to learn invariance to waveform phase inversion so we performed inverse phase augmentation (IPA) by inverting the phase of the training, testing, and validation datasets and concatenating the unchanged datasets and the phase-inverted datasets.
This resulted in 133.5, 106.8, and 26.7 hours of available training, testing, and validation data, respectively.
In contrast, we did not seek to train any level invariance because level normalization is easily accomplished external to the network.

We generated two sets of training/testing/validation corpora: $\mathcal{S}_1$ and $\mathcal{S}_2$.
$\mathcal{S}_1$ used the above process on source datasets 1--8 and reserved the entirety of datasets 9 ($\mathcal{U}_1$) and 10 ($\mathcal{U}_2$) as fully unseen testing data to evaluate how well the network would generalize.
$\mathcal{S}_2$ used the above process on all source datasets thus maximizing the breadth of training but leaving no unseen data.

We used affine transformations to map PESQ and POLQA values ($[1, 4.5]$) and STOI values ($[0, 1]$) to $[-1, 1]$ before use as targets.
For PESQ and POLQA we used a typical range mapping technique.
Since STOI values in our dataset occupy a small portion of the full range possible for STOI output (seen in Fig. \ref{figure:histograms}), we subtracted the mean and divided by the standard deviation.

NAWEnet was trained using mini-batches that were as large as GPU memory would allow, in this case 55 segments per batch.
We used the Adam optimizer \cite{DBLP:journals/corr/KingmaB14} with $10^{-4}$ learning rate, and $L_2$ regularization parameter set to $10^{-5}$.
When the network had trained for an entire epoch we evaluated the validation set and logged the epoch RMSE (root mean-squared error) loss $E_l$ and per-segment correlation between the target and the NAWEnet output, $\rho_{seg}$.
In the case that $E_l$ on the validation set had not decreased by at least $10^{-4}$ for 5 epochs, we multiplied the learning rate by $10^{-1}$.
The network was trained for 30 epochs.

We performed this training process using the NAWEnet architecture for PESQ, POLQA, and STOI targets separately using both the $\mathcal{S}_1$ and $\mathcal{S}_2$ training/testing/validation corpora.
This required a total of six different training sessions and produced six different instances of NAWEnet.
We used PyTorch to construct our datasets, and to construct, train, and test our model. The model was trained on an NVIDIA GeForce GTX 1070.\footnote{Certain products are mentioned in this paper to describe the experiment design. The mention of such entities should not be construed as any endorsement, approval, recommendation, prediction of success, or that they are in any way superior to or more noteworthy than similar entities that were not mentioned.}

\section{Results}
\label{sec:results}

\begin{figure}[!t]
  \includegraphics[width=3.5in]{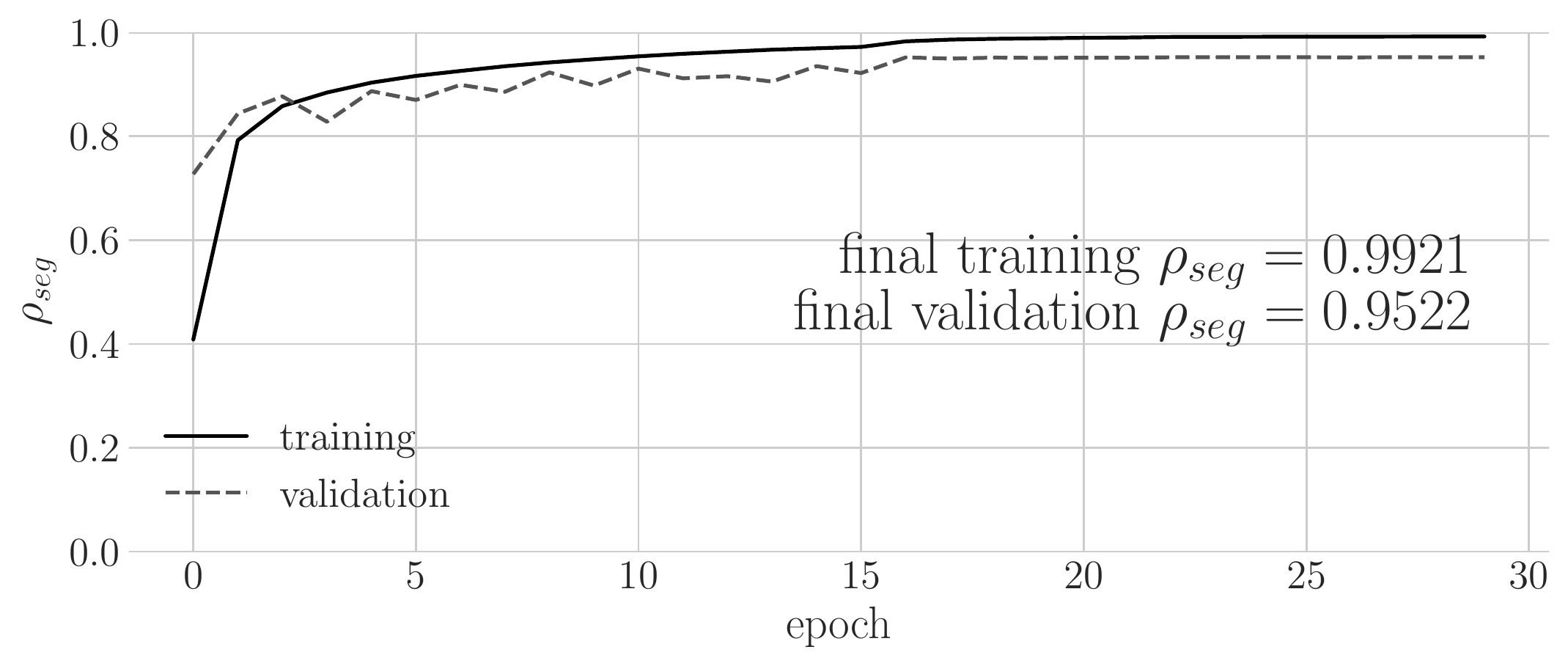}
  \caption{PESQ $\rho_{seg}$ on training and validation portions of $\mathcal{S}_2$ over 30 epochs. A learning rate adjustment occurred shortly after 15 epochs and validation $\rho_{seg}$ then remained fairly constant.\label{figure:train_log}}
\end{figure}

NAWEnet has roughly 40 million parameters to train; about 7 million reside in the convolutional feature extractor and nearly 33 million parameters reside in the first dense layer alone.
It takes about 16 hours to train for 30 epochs on $\mathcal{S}_2$ when mini-batch size is 55 segments.
Table \ref{table:corr_rmse} shows the per-segment correlation $\rho_{seg}$ and RMSE values achieved on the test portion of source datasets individually and combined for all three target metrics.
Fig. \ref{figure:correlation} shows two-dimensional histograms for per-segment target and predicted values, for all three metrics.
The architecture we describe is able to emulate PESQ, POLQA, and STOI with per-segment correlation of 0.95, 0.92, and 0.95 respectively.
Correlations for training data exceed 0.96 in all cases.
Fig. \ref{figure:train_log} is a graph of the training and validation PESQ prediction $\rho_{seg}$ values over the course of 30 epochs of training and demonstrates fast and stable training.
Despite the extreme imbalance in the distribution of POLQA scores demonstrated in Fig. \ref{figure:histograms} (43\% of data is above 4; 14\% above 4.45), NAWEnet manages to achieve $\rho_{seg}>0.91$, a state-of-the art result.
Note that RMSE values for STOI cannot be directly compared with those of PESQ and POLQA because STOI values have a range of 1.0; PESQ and POLQA a range of 3.5

\begin{table*}[h!]
\resizebox{\textwidth}{!}{%
\begin{tabular}{cccccccccccccc}
                                  &                        &                                      & 1                              & 2                              & 3                              & 4                              & 5                              & 6                              & 7                              & 8                              & 9                              & 10                             & combined                       \\
                                  &                        &                                       & $\bullet$                      & $\bullet$                      & $\bullet$                      & $\bullet$                      & $\bullet$                      & $\bullet$                      & $\bullet$                      & $\bullet$                      & $\mathcal{U}_1$                & $\mathcal{U}_2$                & $\mathcal{S}_1$                \\
                                  & PESQ                   & \cellcolor[HTML]{ECF4FF}$\rho_{seg}$ & \cellcolor[HTML]{ECF4FF}0.946 & \cellcolor[HTML]{ECF4FF}0.849 & \cellcolor[HTML]{ECF4FF}0.865 & \cellcolor[HTML]{ECF4FF}0.953 & \cellcolor[HTML]{ECF4FF}0.975 & \cellcolor[HTML]{ECF4FF}0.965 & \cellcolor[HTML]{ECF4FF}0.937 & \cellcolor[HTML]{ECF4FF}0.958 & \cellcolor[HTML]{ECF4FF}0.879 & \cellcolor[HTML]{ECF4FF}0.856 & \cellcolor[HTML]{ECF4FF}0.955 \\
                                  & $T = 13.8\textrm{h}$ & \cellcolor[HTML]{DAE8FC}RMSE         & \cellcolor[HTML]{DAE8FC}0.263 & \cellcolor[HTML]{DAE8FC}0.271 & \cellcolor[HTML]{DAE8FC}0.400 & \cellcolor[HTML]{DAE8FC}0.302 & \cellcolor[HTML]{DAE8FC}0.207 & \cellcolor[HTML]{DAE8FC}0.248 & \cellcolor[HTML]{DAE8FC}0.198 & \cellcolor[HTML]{DAE8FC}0.252 & \cellcolor[HTML]{DAE8FC}0.279 & \cellcolor[HTML]{DAE8FC}0.339 & \cellcolor[HTML]{DAE8FC}0.237 \\
                                  & POLQA                  & \cellcolor[HTML]{DFFFCF}$\rho_{seg}$ & \cellcolor[HTML]{DFFFCF}0.920 & \cellcolor[HTML]{DFFFCF}0.808 & \cellcolor[HTML]{DFFFCF}0.788 & \cellcolor[HTML]{DFFFCF}0.954 & \cellcolor[HTML]{DFFFCF}0.970 & \cellcolor[HTML]{DFFFCF}0.969 & \cellcolor[HTML]{DFFFCF}0.874 & \cellcolor[HTML]{DFFFCF}0.923 & \cellcolor[HTML]{DFFFCF}0.815 & \cellcolor[HTML]{DFFFCF}0.792 & \cellcolor[HTML]{DFFFCF}0.921 \\
                                  & $T = 13.7\textrm{h}$ & \cellcolor[HTML]{D0F0C0}RMSE         & \cellcolor[HTML]{D0F0C0}0.358 & \cellcolor[HTML]{D0F0C0}0.266 & \cellcolor[HTML]{D0F0C0}0.417 & \cellcolor[HTML]{D0F0C0}0.302 & \cellcolor[HTML]{D0F0C0}0.240 & \cellcolor[HTML]{D0F0C0}0.247 & \cellcolor[HTML]{D0F0C0}0.254 & \cellcolor[HTML]{D0F0C0}0.321 & \cellcolor[HTML]{D0F0C0}0.300 & \cellcolor[HTML]{D0F0C0}0.367 & \cellcolor[HTML]{D0F0C0}0.298 \\
                                  & STOI                   & \cellcolor[HTML]{EFEFEF}$\rho_{seg}$       & \cellcolor[HTML]{EFEFEF}0.942 & \cellcolor[HTML]{EFEFEF}0.886 & \cellcolor[HTML]{EFEFEF}0.895 & \cellcolor[HTML]{EFEFEF}0.961 & \cellcolor[HTML]{EFEFEF}0.953 & \cellcolor[HTML]{EFEFEF}0.952 & \cellcolor[HTML]{EFEFEF}0.941 & \cellcolor[HTML]{EFEFEF}0.950 & \cellcolor[HTML]{EFEFEF}0.774 & \cellcolor[HTML]{EFEFEF}0.809 & \cellcolor[HTML]{EFEFEF}0.947 \\
\multirow{-7}{*}{$\mathcal{S}_1$} & $T = 13.6\textrm{h}$ & \cellcolor[HTML]{DFDFDF}RMSE         & \cellcolor[HTML]{DFDFDF}0.025 & \cellcolor[HTML]{DFDFDF}0.016 & \cellcolor[HTML]{DFDFDF}0.034 & \cellcolor[HTML]{DFDFDF}0.020 & \cellcolor[HTML]{DFDFDF}0.028 & \cellcolor[HTML]{DFDFDF}0.023 & \cellcolor[HTML]{DFDFDF}0.019 & \cellcolor[HTML]{DFDFDF}0.026 & \cellcolor[HTML]{DFDFDF}0.022 & \cellcolor[HTML]{DFDFDF}0.026 & \cellcolor[HTML]{DFDFDF}0.024 \\ \hline\hline
\\
\hline\hline
                                  &                        &                                      & 1                              & 2                              & 3                              & 4                              & 5                              & 6                              & 7                              & 8                              & 9                              & 10                             & combined                       \\
                                  &                        &                                 & $\bullet$                      & $\bullet$                      & $\bullet$                      & $\bullet$                      & $\bullet$                      & $\bullet$                      & $\bullet$                      & $\bullet$                      & $\bullet$                      & $\bullet$                      & $\mathcal{S}_2$                \\
                                  & PESQ                   & \cellcolor[HTML]{ECF4FF}$\rho_{seg}$ & \cellcolor[HTML]{ECF4FF}0.950 & \cellcolor[HTML]{ECF4FF}0.880 & \cellcolor[HTML]{ECF4FF}0.877 & \cellcolor[HTML]{ECF4FF}0.955 & \cellcolor[HTML]{ECF4FF}0.969 & \cellcolor[HTML]{ECF4FF}0.974 & \cellcolor[HTML]{ECF4FF}0.938 & \cellcolor[HTML]{ECF4FF}0.961 & \cellcolor[HTML]{ECF4FF}0.915 & \cellcolor[HTML]{ECF4FF}0.889 & \cellcolor[HTML]{ECF4FF}0.953 \\
                                  & $T = 16.5\textrm{h}$ & \cellcolor[HTML]{DAE8FC}RMSE         & \cellcolor[HTML]{DAE8FC}0.252 & \cellcolor[HTML]{DAE8FC}0.251 & \cellcolor[HTML]{DAE8FC}0.383 & \cellcolor[HTML]{DAE8FC}0.285 & \cellcolor[HTML]{DAE8FC}0.222 & \cellcolor[HTML]{DAE8FC}0.229 & \cellcolor[HTML]{DAE8FC}0.197 & \cellcolor[HTML]{DAE8FC}0.242 & \cellcolor[HTML]{DAE8FC}0.236 & \cellcolor[HTML]{DAE8FC}0.293 & \cellcolor[HTML]{DAE8FC}0.236 \\
                                  & POLQA                  & \cellcolor[HTML]{DFFFCF}$\rho_{seg}$ & \cellcolor[HTML]{DFFFCF}0.929 & \cellcolor[HTML]{DFFFCF}0.814 & \cellcolor[HTML]{DFFFCF}0.803 & \cellcolor[HTML]{DFFFCF}0.963 & \cellcolor[HTML]{DFFFCF}0.973 & \cellcolor[HTML]{DFFFCF}0.962 & \cellcolor[HTML]{DFFFCF}0.870 & \cellcolor[HTML]{DFFFCF}0.923 & \cellcolor[HTML]{DFFFCF}0.855 & \cellcolor[HTML]{DFFFCF}0.827 & \cellcolor[HTML]{DFFFCF}0.915 \\
                                  & $T = 16.1\textrm{h}$ & \cellcolor[HTML]{D0F0C0}RMSE         & \cellcolor[HTML]{D0F0C0}0.336 & \cellcolor[HTML]{D0F0C0}0.283 & \cellcolor[HTML]{D0F0C0}0.433 & \cellcolor[HTML]{D0F0C0}0.281 & \cellcolor[HTML]{D0F0C0}0.225 & \cellcolor[HTML]{D0F0C0}0.267 & \cellcolor[HTML]{D0F0C0}0.260 & \cellcolor[HTML]{D0F0C0}0.320 & \cellcolor[HTML]{D0F0C0}0.242 & \cellcolor[HTML]{D0F0C0}0.335 & \cellcolor[HTML]{D0F0C0}0.297 \\
                                  & STOI                   & \cellcolor[HTML]{EFEFEF}$\rho_{seg}$ & \cellcolor[HTML]{EFEFEF}0.935 & \cellcolor[HTML]{EFEFEF}0.891 & \cellcolor[HTML]{EFEFEF}0.881 & \cellcolor[HTML]{EFEFEF}0.964 & \cellcolor[HTML]{EFEFEF}0.955 & \cellcolor[HTML]{EFEFEF}0.958 & \cellcolor[HTML]{EFEFEF}0.943 & \cellcolor[HTML]{EFEFEF}0.953 & \cellcolor[HTML]{EFEFEF}0.863 & \cellcolor[HTML]{EFEFEF}0.873 & \cellcolor[HTML]{EFEFEF}0.946 \\
\multirow{-7}{*}{$\mathcal{S}_2$} & $T = 16\textrm{h}$   & \cellcolor[HTML]{DFDFDF}RMSE         & \cellcolor[HTML]{DFDFDF}0.024 & \cellcolor[HTML]{DFDFDF}0.016 & \cellcolor[HTML]{DFDFDF}0.035 & \cellcolor[HTML]{DFDFDF}0.020 & \cellcolor[HTML]{DFDFDF}0.026 & \cellcolor[HTML]{DFDFDF}0.022 & \cellcolor[HTML]{DFDFDF}0.019 & \cellcolor[HTML]{DFDFDF}0.024 & \cellcolor[HTML]{DFDFDF}0.016 & \cellcolor[HTML]{DFDFDF}0.022 & \cellcolor[HTML]{DFDFDF}0.022
\end{tabular}%
}
\caption{Per-segment Pearson correlation and RMSE achieved on testing data after training NAWEnet to target PESQ, POLQA and STOI separately. A $\bullet$ indicates the source dataset was included in the training process. Results in the ``combined'' column reflect evaluation on the test portion of the specified aggregated dataset; $\mathcal{S}_1$ or $\mathcal{S}_2$. Results in columns corresponding to $\mathcal{U}_1$ and $\mathcal{U}_2$ in $\mathcal{S}_1$ reflect evaluation on the entirety of datasets 9 and 10 separately as completely unseen data. Training time reported as $T$. MOS values are available for source datasets 1-6; PESQ and POLQA (FR) have combined $\rho_{seg}=0.81$; ANIQUE+ and P.563 (NR) have combined $\rho_{seg}=0.60$ and $\rho_{seg}=0.53$ respectively.}
\label{table:corr_rmse}
\end{table*}

To put these results in context we can compare FR quality metrics PESQ and POLQA with the MOS scores that are available for source datasets 1-6.
The correlation between these MOS scores and either PESQ or POLQA is $\rho_{seg}=0.81$.
The NR quality metrics ANIQUE+ and P.563 achieve $\rho_{seg}=0.60$ and $0.53$ respectively.
With $\rho_{seg}$ above 0.91, NAWEnet results agree with FR measures better than those FR measures agree with MOS and far better than other NR measures agree with MOS.
Though we have not trained NAWEnet to learn a MOS target directly, we have demonstrated the framework's flexibility to learn very different targets given sufficient data.

\vspace{3ex}

In contrast to recent work, NAWEnet accepts the waveform itself as input, learns appropriate features, and predicts one of three target metrics.
Quality-Net \cite{Fu} targets only PESQ and uses magnitude spectrum as an input but has the advantage of not requiring a fixed-length input.
The best results in \cite{Avila2019} ($\rho=0.87$) target only crowd-MOS and were achieved by using Mel-cepstral coefficients and other derived features as an input to a dense network.
NISQA \cite{Mittag2019} achieves a per-condition correlation of 0.89, targeting MOS and POLQA jointly, by first creating a set of spectrograms and then further processing them with a Mel filter bank.
The authors of \cite{Ooster2019} calculate features for input to a DNN and further process the output while targeting only MOS.

%\vspace{3ex}

NAWEnet instances tasked with learning PESQ and POLQA show greater $\rho_{seg}$ on source dataset 8 than on source dataset 7.
The two source datasets share common speech source but dataset 8 includes frame erasures and concealments.
The higher correlation on dataset 8 is unexpected because frame erasure and concealment typically makes quality measurement more difficult.
A possible explanation is that dataset 8 is large---it constitutes 46\% of $\mathcal{S}_2$.

Examining source datasets 9 and 10 in $\mathcal{S}_1$ we see that NAWEnet had some difficulty generalizing to completely unseen data with $\rho_{seg}$ roughly equivalent to the two more difficult source datasets: 2 and 3.
In addition to having 21 unseen talkers, source datasets 9 and 10 also had the lowest average speech activity of all source datasets.
However, when source datasets 9 and 10 are included as part of $\mathcal{S}_2$, $\rho_{seg}$ on the test portions of those datasets improves.
Source datasets 9 and 10 constitute 15\% of data in $\mathcal{S}_2$.
This shows that NAWEnet is capable of learning to handle lower levels of speech activity and new talkers with commensurate training data.
Values for $\rho_{seg}$ in the test portions of source datasets 1-8 for $\mathcal{S}_2$ improved in 17 of 24 cases (8 source datasets $\times$ 3 target measurements) compared to $\mathcal{S}_1$ but were not significantly harmed otherwise. 
This indicates that the NAWEnet architecture could improve performance on unseen inputs with a carefully tuned training corpus.

By reducing training time and improving accuracy the PReLU activation function was found to have superior performance to the popular leaky ReLU activation.
Allowing the network to learn $f_n$ or $d_o$ distinct PReLU parameters per section significantly increases the flexibility of the network without adding an undue number of parameters.

Because $S_1$ in the convolutional feature extractor is operating on raw audio samples it is slightly different than the rest of the convolutional sections.
It was experimentally found that $f_l=11$ performed better than $f_l=7$ in the $S_1$, but $f_l>11$ gave no additional benefit.
The superiority of the slightly longer and more selective filter is consistent with the intuitive notion that emphasizing or attenuating specific frequencies in the first layer is an important step towards feature extraction.
We found that average pooling is superior to max-pooling for downsampling the first layer and this is consistent with the observation that no single audio sample is more important than the next.

\cheat
\section{Conclusion}
\label{sec:conclusion}
We have demonstrated that the NAWEnet design is flexible and can quickly learn the necessary features and mappings to emulate two different NB speech quality metrics and a speech intelligibility metric.

Future work includes testing suitability of these new networks for transfer learning, pruning the number of parameters in the dense portion of the network, using more sophisticated training techniques, and implementing additional deep learning best practices.
It may be beneficial to teach the networks to learn auto-regressive moving-average filters rather than simple moving average filters.
We plan to inspect our results to see if it is possible to know what features NAWEnet is learning and how those features are being quantified and combined to produce speech quality or speech intelligibility values.
We also plan to use the WEnets framework to address higher sample rates.
Though it is difficult to find a subjective quality database that is large enough to train a convolutional network, it would be very interesting to see how this framework performs on subjective test scores.

\balance
\bibliographystyle{IEEEbib}

\end{document}